\documentclass[preprint,pra,aps,showpacs,preprintnumbers,amsmath,amssymb,eqsecnum]{revtex4}

\usepackage{graphicx}
\usepackage{dcolumn}
\usepackage{bm}

\def\E{{\cal E}}
\def\x{{\bf x}}
\def\y{\mathbf{y}}
\def\k{\mathbf{k}}
\def\q{\mathbf{q}}

\def\r{\mathbf{r}}

\def\b{\beta}

\def\ih{{ \frac{i}{\hbar} }}

\def\half{\frac{1}{2}}
\def\ria{{\rightarrow}}

\def\E{{\cal E}}
\def\x{{\bf x}}
\def\y{\mathbf{y}}
\def\k{\mathbf{k}}
\def\q{\mathbf{q}}

\def\r{\mathbf{r}}

\def\b{\beta}

\def\ih{{ \frac{i}{\hbar} }}

\def\half{\frac{1}{2}}

\newcommand\beq{\begin{equation}}
\newcommand\eeq{\end{equation}}
\newcommand\bea{\begin{eqnarray}}
\newcommand\eea {\end{eqnarray}}

\begin{document}


\title{Two Derivations of the Master Equation \\ of Quantum Brownian Motion}

\author{J.J.Halliwell}%
\affiliation{Blackett Laboratory \\ Imperial College \\ London SW7
2BZ \\ UK }



\begin{abstract}
Central to many discussion of decoherence is a master equation for
the reduced density matrix of a massive particle experiencing
scattering from its surrounding environment, such as that of Joos
and Zeh. Such master equations enjoy a close relationship with spontaneous localization
models, like the GRW model.
This aim of this paper is to present two derivations of
the master equation. The
first derivation is a pedagogical model designed to illustrate the
origins of the master equation as simply as possible, focusing on
physical principles and without the complications of $S$-matrix
theory. This derivation may serve as a useful tutorial
example for students attempting to learn this subject area. The
second is the opposite: a very general derivation using
non-relativistic many body field theory. It reduces to the
equation of the type given by Joos and Zeh in the one-particle
sector, but correcting certain numerical factors which have
recently become significant in connection with experimental tests
of decoherence. This master equation also emphasizes the role of
local number density as the ``preferred basis'' for decoherence in
this model.

\end{abstract}

\maketitle

\section{Introduction}

Non-unitary master equations for a density matrix arise in both
continuous state localization models, such as GRW theory \cite{GRW}, and
in decoherence calculations in standard quantum theory. They
differ, however, in their underlying physical pictures.
GRW theory involves an explicitly modified dynamics in the
Schr\"odinger equation. Standard decoherence calculations,
by contrast, employ the usual Schr\"odinger dynamics for
a system coupled to its environment, but this becomes a non-unitary
dynamics for the reduced density matrix once the environment
is traced out. These similarities and differences have undoubtedly
been clear ever since the appearance of GRW theory twenty years ago,
but there is surely still more to learn in this area. It is a great
pleasure to have the opportunity to contribute to this volume
in honour of the 70th birthday of GianCarlo Ghirardi.

This paper concerns master equations in standard quantum mechanics
of the type used in decoherence studies, obtained by coupling
a point particle by some kind of environment. A simple example
of such a master equation in one dimension is
\begin{eqnarray}
{\partial \rho \over \partial t} & = &
{i \hbar \over 2 M} \left( { \partial^2 \rho \over \partial x^2}
- {\partial^2 \rho \over \partial y^2} \right)
- D (x-y)^2 \rho
\label{1.1}
\end{eqnarray}
This describes quantum Brownian motion for a free particle of mass $M$
in the limit
of negligible dissipation.
This is the one-dimensional version of an equation first obtained
by Joos and Zeh \cite{JoZ} (in the limit of small $|x-y|)$,
which involves a massive free particle undergoing
scatterings by an environment of much lighter particles. Many other
derivations of this and similar equations have since been given
\cite{GaF,Dio,Vac,HoSi,Omn1}.
This equation can also arise for the case of a point particle linearly
coupled to a thermal bath of harmonic oscillators, and this model also
has been the subject of many papers \cite{CaL,HPZ,HaYu}. Models
involving fields have also been considered \cite{HaDo,UnZu,AnZo}.
(The literature on quantum Brownian motion is considerable so only
a selection is mentioned here. See also Ref.\cite{QBM}.)

A key reference point in these studies is
the Lindblad form of
the master equation \cite{Lin}, which is the most general possible form a
master equation can take under the assumption that the evolution
is Markovian (a condition well-satisfied in a wide variety of
interesting models). The Lindblad master equation is
\begin{equation}
\frac { d \rho}{ dt} = -i  [H, \rho] - \half \sum_{j=1}^n \left( \{
L_j^{\dag} L_j, \rho \} - 2 L_j \rho L_j^{\dag} \right)
\label{1.2}
\end{equation}
Here, $H$ is the Hamiltonian of the distinguished subsystem
(sometimes modified by terms depending on the $L_j$) and the $n$
operators $L_j$ model the effects of the environment. For example, the master
equation of one-dimensional quantum Brownian motion, including
dissipation, is of the Lindblad form with a single Lindblad operator
\begin{equation}
L = \left( \frac {4 M \gamma k T } { \hbar^2} \right)^{\half} x
+ i \left( \frac {\gamma}
{2 M k T } \right)^{\half} p
\label{1.3}
\end{equation}
as described in Refs.\cite{Dio2,HaZ}. (This reproduces Eq.(\ref{1.1})
for small $\gamma$ with $D = 2 M \gamma k T / \hbar^2$).

Equation (\ref{1.1}) describes the decoherence process in which an arbitrary
initial density matrix becomes approximately diagonal in position
on a very short timescale. This process is thought to be
a key element in understanding how classical behaviour emerges from
quantum theory \cite{Har6,Zur1,Hal0}. Recent experiments have been able to actually observe
the rate of the decoherence process \cite{Exp}, which is connected to the constant
$D$ in Eq.(\ref{1.1}). It turns out that some of the original derivations of
the master equation led to incorrect values of $D$. More recent derivations
\cite{HoSi,HaDo} have corrected these errors and produce values of $D$ compatible with
experiments. (See also Refs.\cite{Hor,Adl}).

The purpose of this paper is to present two derivations of a class
of master equations of the form Eq.(\ref{1.1}) and its generalizations.

The first derivation, described in Section 2 and 3, is a simple pedagogical model, designed
to illustrate the way in which the general form of Eq.(\ref{1.1}) follows from
some simple physical ideas. We therefore avoid the technical complications
and sometimes non-transparent mathematical assumptions involved in these
models, but make no claims about making physically accurate predictions.
(This model is similar to one presented by Joos et al. \cite{Joos}).

The second derivation, described in Sections 4, 5 and 6,
is a very general derivation
of a class of master equations and makes use of non-relativistic many body field theory.
It reduces, in form, to an equation of the type given by Joos and Zeh and others,
but with a correct value of the decoherence rate that is compatible with experiments.
This derivation is also relevant to another issue in decoherence theory, which is
the question of the ``preferred basis'' -- the natural basis in which
interferences are destroyed. In the master equation Eq.(\ref{1.1}), it is clearly
the interferences between different values of {\it position} that are initially suppressed,
so that position is the preferred basis. More generally, it is known that
under evolution according to the Lindblad form Eq.(\ref{1.2}), it is the Lindblad
operators $L_j$ that defined the preferred basis (at least in simple models) \cite{DGHP}.
We will see in the many body field theory
model that most generally in system-environment models, {\it local number density}
is the preferred basis (with position emerging as
a special case of this in the one-particle sector). The special
role of number density also echoes certain aspects of the GRW model \cite{GRW}. Furthermore,
local densities are thought to play a key role in the most general possible
derivations of emergent classicality, even when there is no environment
present \cite{Hal0,HalHydro}

The many body field theory derivation of the master equation
first appeared, in essence, in Ref.\cite{HaDo},
as part of a wider investigation into the properties of the decoherent histories
approach to quantum theory \cite{GeH1,GeH2,Har1,Gri,Omn}. Here it is presented on its own merits as a contribution
to the theory of quantum Brownian motion.

\section{A Simple Model}

We consider a system consisting of a particle in one dimension which
interacts through occasional collisions with a gas of light
particles. Except for the collisions, which are assumed to be very
brief, the particles evolve freely. By considering the change in
the system density matrix during these collisions we will derive
the form of the decoherence term (the last term) in Eq.(\ref{1.1}).

We first consider the collision process classically. Suppose the
system particle has momentum $P$ and mass $M$ and a particle from
the environment has momentum $p$ and mass $m$. We will assume that
the collision conserves both energy and momentum. If the final
momenta are $P'$ and $p'$, we therefore have
\begin{eqnarray}
\frac {P^2} {2M} + \frac {p^2} {2m} &=& \frac {{P'}^2} {2M} +
\frac {{p'}^2} {2m}
\\
P + p &=& P' + p'
\end{eqnarray}
Ignoring the trivial solution $P= P'$ and $p=p'$, the final
momenta are given by
\begin{eqnarray}
P' &=& \frac { (M-m) } { (M+m) } P + \frac {2M} { (M+m)} p
\label{2A.3}
\\
&\equiv & a P + b p
\label{2A.4}
\\
p' &=& \frac { 2m } { (M+m) } P - \frac { (M-m) } { (M+m)} p
\label{2A.5}
\\
&\equiv & c P - a p
\label{2A.6}
\end{eqnarray}
(where the coefficients $a,b,c$ are read off from Eqs.(\ref{2A.3}),(\ref{2A.5})).
It is now very useful to make two approximations. We first assume
that the environment particles are much lighter than the system
particles:
\begin{equation}
m \ll M
\end{equation}
Second, we assume that the speed of the system particle is much
smaller than the speed of environment particles,
\begin{equation}
\frac { P } {M }  \ll \frac {p} {m}
\end{equation}
(although the momenta $P$ and $p$ may be comparable in size).
These approximations imply that the momenta after the collision
are given by the much simpler expressions
\begin{eqnarray}
P' & \approx& P + 2 p
\label{2A.9}\\
p' &\approx& -p
\label{2A.10}
\end{eqnarray}

Turn now to the quantum case. We will make the reasonable
assumption that energy and momentum are also conserved by the
quantum description of the collision. The key idea in the quantum
case is to work with states of definite momentum (plane wave
states) and use the above results to deduce how they change
during a collision. We will eventually also need to assume
something about the locality of the interaction, but that will
not be needed just yet.

Introducing the positions $(x,q)$ of the
system and environment particles, it follows from Eqs.(\ref{2A.9}), (\ref{2A.10})
that an initial plane wave for the
total system changes according to
\begin{eqnarray}
e^{ \ih P x } e^ { \ih p q} \ &\ria& \ e^{ \ih (P + 2 p ) x} e^{ -
\ih p q}
\\
&=& \ e^{ \ih P x } e^{ \ih p (2x - q) }
\end{eqnarray}
(in the approximations described above).
Or, in configuration space, the effect of the collision is to make the replacement
\bea
x  & \ria & x
\label{2A.13}
\\
q  & \ria & 2x - q
\label{2A.14}
\eea
Since the plane waves are a complete
set of states, this result determines the effects of a collision on any
initial state.

We suppose that the initial density matrix of the whole system is
a simple product state, $ \rho(x,y) \ \rho_{\E} (q,q') $. As a
result of a single collision, Eqs.(\ref{2A.13}), (\ref{2A.14}) imply that the
total density matrix changes according to
\begin{equation}
\rho (x,y)\  \rho_{\E} (q,q') \ \ria \ \rho( x,y)\ \rho_{\E}
(2x-q, 2y - q')
\end{equation}
Tracing over the environment, this means that the density operator
of the system only evolves according to
\begin{equation}
\rho (x,y) \ \ria \ \rho (x,y) \ \int dq \ \rho_{\E} (2x-q, 2y -
q)
\end{equation}
We suppose that the collisions take place at a rate $\Gamma$ per
unit time. The total change in the density matrix of the system
during a small time interval $\Delta t$ is therefore given by
\begin{equation}
\Delta \rho (x,y) = - \Delta t \ F (x,y) \ \rho (x,y)
\end{equation}
where
\begin{equation}
F(x,y) =  \Gamma \ \left( 1 - \int dq \ \rho_{\E} (2x - q, 2y - q)
\right)
\end{equation}
This is also usefully written,
\begin{equation}
F(x,y) = \Gamma \ {\rm Tr} \left( ( 1 - e^{ 2 \ih (x-y) \hat p } )
\hat \rho_{\E} \right)
\end{equation}
where $\hat p$ is the momentum operator on the environment and the
trace is over a complete set of environment states. It follows
that the master equation is
\begin{equation}
\frac {\partial \rho} {\partial t} = \frac {i \hbar} {2 M}  \left(
\frac {\partial^2 \rho} { \partial x^2} - \frac {\partial^2 \rho}
{\partial y^2} \right) - F(x,y) \rho
\end{equation}

In the small $|x-y|$ limit we have
\begin{equation}
F(x,y) = \frac {2 \Gamma \langle \hat p^2 \rangle }  {\hbar^2}
(x-y)^2
\end{equation}
where we have assumed that the environment state is such that $
\langle \hat p \rangle = 0 $. For a thermal environment state we
have
\begin{equation}
\langle \hat p^2 \rangle = m k T
\end{equation}
so we obtain a master equation of the form Eq.(\ref{1.1}) with $D =
2 m \Gamma k T / \hbar^2$.
We have therefore obtained the expected form of the master equation,
using energy and momentum conservation, together with the approximations
that the environment particles are much lighter and faster than the system
particle.

A slightly different but simpler model along these lines was given
by Joos at al \cite{Joos}. Their model postulates a simple dynamics
involving a particle being subject to random kicks, but without relating it to
a collision process with energy and momentum conservation as here.

\section{A Wigner Function Derivation Including Dissipation}

It is now useful to
give a more detailed derivation of the master equation in this simple
model using the Wigner representation. We go beyond the approximations used
above and work to leading order in $m/M$.
This derivation shows how the dissipative terms arises and also establishes
the connection between the collision rate $\Gamma $ and the dissipation $ \gamma$,
hence connects microscopic and macroscopic parameters.

The Wigner function of the density matrix $\rho(x,y)$ of a one-dimensional systems
is defined by
\beq
W(p,q) = { 1 \over 2 \pi \hbar} \int d \xi \ e^{-\ih p \xi}
\ \rho( q + \half \xi, q - \half \xi)
\eeq
together with its inverse
\beq
\rho(x,y) = \int dp \ e^{\ih p (x-y) } \ W ( p, {x+y \over 2} )
\eeq
The Wigner function has the properties
\bea
\int dp W (p,q) &=& \rho(q,q)
\label{3A.3}
\\
\int dq W (p,q) &=& \tilde \rho (p,p)
\label{3A.4}
\eea
where $\tilde \rho $ is the Fourier transform of $\rho  (x,y)$,
so $W$ contains the usual position and momentum probabilities
as its marginal distributions.
The Wigner function is the closest thing quantum mechanics has to
a phase space probability distribution function but narrowly fails
since $W$ is not always positive. Its time evolution is identical
to classical evolution for linear systems, with corrections proportional
to powers of $\hbar^2$ for non-linear potentials. It is therefore a very
useful tool for discussing the connection between quantum
and classical systems, although note that there are subtle
differences between the Wigner function and a classical distribution
function, as we shall see shortly. (See Refs.\cite{Wig} for properties
of the Wigner function.)

In the absence of interactions, the two particle Wigner function
$W_2 (P,X,p,q) $ obeys the equation
\begin{equation}
\frac {\partial W_2 } { \partial t} = - \frac { P} {M} \frac
{\partial W_2} {\partial X} - \frac { p } { m } \frac {\partial
W_2 } { \partial q}
\end{equation}
and this corresponds to unitary evolution of the density matrix.
We need to find terms representing the collision to add to the right-hand side,
using an argument along the lines of that used in the previous section for the density matrix.
We again consider the collision process described by Eqs.(\ref{2A.4}), (\ref{2A.6})
and look for the change in the Wigner function.

If $W$ were a classical distribution
function, it would be sufficient to consider only the momentum transfer
described by Eqs.(\ref{2A.4}), (\ref{2A.6}). In the Wigner function, however,
things are a bit more subtle. If Eqs.(\ref{2A.4}), (\ref{2A.6})
describe a process in the momentum representation of a quantum system,
then there is a corresponding transformation in position space.
Proceeding along lines identical to the derivation of Eqs.(\ref{2A.13}),
(\ref{2A.14}), it is easy to see that this transformation is
\bea
X & \ria & a X + c q = X + c (q-X)
\label{3A.6}
\\
q & \ria & b X - a q = 2 X - q + c( q - X)
\label{3A.7}
\eea
That is, the collision process in the Wigner function involves {\it both}
the transformation Eqs.(\ref{2A.4}), (\ref{2A.6}) on the momenta together
with the transformation Eqs.(\ref{3A.6}), (\ref{3A.7}) on the positions.
This must be the case because on integrating out the
momenta in the Wigner function, the correct distributions
for position must be obtained, and vice versa (as in Eqs.(\ref{3A.3}), (\ref{3A.4})).

Denoting the system Wigner function by $W(P,X)$ and the
environment Wigner function by $W_{\E} (p,q)$, the above discussion
implies that the effect of a
collision is to produce the transition
\begin{equation}
W (P,X)\ W_{\E} (p,q) \ \rightarrow \ W (aP + b p, a X  + cq ) \ W_{\E} (
c P - ap , b X - a q)
\label{3A.8}
\end{equation}
in the two particle Wigner function. However, in this derivation it
turns out that it is also important to incorporate the fact that
the interaction is local, that is, it is described
by a potential of the form $ V (x-q)$ which decays for large $ | x - q |$. In a
more complete derivation this would be accomplished by multiplying Eq.(\ref{3A.8}) by a function
of $X - q $ which is concentrated around $ X = q$. However, in the interests of keeping
the derivation heuristic,
we will incorporate this ``by hand'', by assuming that terms of the form
$ | q - X | $ are small in some sense. Since the constant $c$ is also small
(of order $m/M$), the most minimal and natural
implementation of locality is to simply drop the terms $ c ( q - X)$ in
Eqs.(\ref{3A.6}), (\ref{3A.7}).

Integrating out the
environment, and considering $\Gamma$ collisions per unit time, we now have that
the evolution equation of the system Wigner function is
\begin{equation}
\frac {\partial W } { \partial t} = - \frac { P} {M} \frac
{\partial W} {\partial X} + \Gamma \int dp dq \left[ \ W (aP + b
p, X ) \ W_{\E} ( c P - ap , 2 X - q) - W (P,X) \ W_{\E} (p,q) \right]
\label{3A.9}
\end{equation}
which is a Boltzmann equation.
Using some simple changes of
variable for $p$ and $q$ in the first term, this is easily rewritten,
\begin{equation}
\frac {\partial W } { \partial t} = - \frac { P} {M} \frac
{\partial W} {\partial X} + \Gamma \int dp dq \left[ \ \frac {1}
{a} W ( \frac{P} {a}  - \frac{b }{a} p , X ) - W (P,X) \right]
W_{\E} (p,q)
\label{3A.10}
\end{equation}
(From this form we see that the position coordinates in the Wigner
function are, in the end, effectively unchanged as a result of the collision,
despite the argument above, but this
is due to the approximation of dropping terms proportional to
$ c (q - X)$).

We again assume that the environment is in a thermal state at
temperature $T$. To obtain the more familiar form of the equation,
we will assume that $W$ is slowly varying in $P$ so that it may be
expanded in derivatives. We will also take $m \ll M$ and look for
the leading order terms in $ m / M $, so we have that
\begin{equation}
\frac {1} {a} = \frac {M+m} {M-m} \approx 1 + \frac {2 m } { M }
\end{equation}
We thus obtain
\begin{equation}
\frac {\partial W } { \partial t} = - \frac { P} {M} \frac
{\partial W} {\partial X} + \Gamma \left( \frac {2m } {M} W +
\frac {2 m} {M} P \frac {\partial W} { \partial P} + \frac {1} {2}
\left( \frac {b} {a} \right)^2 \langle p^2 \rangle \frac
{\partial^2 W} {\partial P^2} \right)
\end{equation}
where we have dropped terms of order $ m^2 / M^2 $ and terms
involving higher derivatives of $W$. Now we note that it is
appropriate to identify the dissipation $\gamma$ as
\begin{equation}
\gamma = \frac {  m } {M } \Gamma
\end{equation}
hence we make a connection between the collision rate and the
dissipation. Since $b/a \approx 2$ to leading order and $\langle
p^2 \rangle = 2 m k T $, we finally obtain the result
\begin{equation}
\frac {\partial W } { \partial t} = - \frac { P} {M} \frac
{\partial W} {\partial X} +
 2 \gamma \frac {\partial (P W) } { \partial P}
+ 2 M \gamma k T  \frac {\partial^2 W} {\partial P^2}
\label{3A.14}
\end{equation}
This is the expected Wigner equation for a system undergoing
quantum Brownian motion including dissipation.
As is well known, this equation describes the approach to thermal
equilibrium. Inverting the Wigner transform, the corresponding density matrix equation is
\beq
{\partial \rho \over \partial t}  =
{i \hbar \over 2 M} \left( { \partial^2 \rho \over \partial x^2}
- {\partial^2 \rho \over \partial y^2} \right)
- i \hbar \gamma (x-y) \left( { \partial \rho \over \partial x} -
{ \partial \rho \over \partial y} \right)
- \frac {2 M \gamma k T} {\hbar^2} (x-y)^2 \rho
\label{3A.15}
\eeq
Although often given as the master equation for quantum Brownian motion
with dissipation \cite{CaL}, this equation is not in fact of the Lindblad
form Eq.(\ref{1.2}) and actually suffers from a possible small positivity
violation \cite{Amb,AnHa}. However, it can easily be modified into the Lindblad
form Eq.(\ref{1.2}) with Lindblad operator
Eq.(\ref{1.3}) by addition of a term with coefficient
proportional to $ 1 / T $, so the difference is neglible
for high temperatures.

In summary, we obtain the master equation Eq.(\ref{3A.9}) with dissipation using
momentum and energy conservation to describe the collision, and taking
leading and first orders terms in order $ m / M $ together with
a simple approximation to incorporate the the locality
of the interaction. The familiar forms Eqs.(\ref{3A.14}), (\ref{3A.15}) are obtained using
the further assumption that the Wigner function is a slowly varying
function of $P$ (which, in the density matrix, corresponds to exploring the
region of small $ | x-y| $).

\section{Many Body Field Theory}

We now turn to the second derivation of the master equation, using
many body field theory. This section is based on Ref.\cite{HaDo}.
We begin by briefly reviewing the formalism
\cite{Zub,FeWa}
We consider a set of non-relativistic system particles described by a field
$\psi (\x)$ interacting through a potential $\phi (\x)$ with an environment
described by a field $\chi (\x) $. The total system is described by the Hamiltonian
\begin{eqnarray}
H &=& \int d^3 x \ \left( { 1  \over 2M} \nabla \psi^{\dag} (\x) \cdot \nabla \psi (\x)
+  { 1 \over 2m} \nabla \chi^{\dag} (\x) \cdot \nabla \chi (\x) \right)
\nonumber \\
&+& \half \int d^3x d^3 x' \ \psi^{\dag} (\x ) \psi (\x') \phi (\x
- \x') \chi^{\dag} (\x') \chi (\x)
\label{2.1}
\end{eqnarray}
(For simplicity we set $\hbar = 1$ hereafter).
In this language, the number densities $N(\x )$ and $n(\x )$ of the system and environment
fields are
\begin{eqnarray}
N (\x ) &=& \psi^{\dag} (\x) \psi (\x)
\\
n (\x ) &=& \chi^{\dag} (\x) \chi (\x)
\label{2.2}
\end{eqnarray}

The above relations are also more conveniently written in terms of
$a_{\k}$ and $b_{\k}$, the annihilation operators for
the system and environment, respectively, and the Hamiltonian then is
\begin{eqnarray}
H &=& \sum_\q \left( E_{\q} a^{\dag}_{\q} a_{\q}
+ \omega_{\q} b^{\dag}_{\q} b_{\q} \right)
\\\nonumber
&+& {1 \over 2 V} \sum_{\k_1' + \k_2' = \k_1 + \k_2}
\nu ( \k_2' - \k_2 ) a^{\dag}_{\k_1} b^{\dag}_{\k_2}
a_{\k_1'} b_{\k_2'}
\label{2.3}
\end{eqnarray}
where $ E_{\q} = \q^2 / 2 M $, $\omega_{\q} = \q^2 / 2 m $,
$V$ is the spatial volume of the system (which we assume is in a box) and
\begin{equation}
\nu (\k) = \int d^3x \ e^{ - i \k \cdot \x} \ \phi (\x)
\end{equation}
The Fourier
transformed number densities are
\begin{eqnarray}
N_\k  &=& \sum_{\q} a^{\dag}_{\q} a_{\q + \k}
\\
n_\k  &=& \sum_{\q} b^{\dag}_{\q} b_{\q + \k}
\label{2.4}
\end{eqnarray}
and one may see that the Hamiltonian has the more concise form
\begin{eqnarray}
H & = &\sum_\q \left( E_{\q} a^{\dag}_{\q} a_{\q}
+  \omega_{\q} b^{\dag}_{\q} b_{\q} \right)
+ {1 \over 2 V} \sum_{\k} \nu (\k) N_{\k} n_{-\k}
\\
&=& H_0 + H_{int}
\label{2.5}
\end{eqnarray}
From these relations we see that the environment couples to the number density of the
system. It is this feature of many body field theory that makes it the appropriate
medium for the derivation of the master equation emphasizing the role of number density.

The $S$-matrix is
\begin{equation}
S = T \exp \left( - i \int_{ - \infty}^{\infty} dt \ H_{int} (t) \right)
\label{2.6}
\end{equation}
where
\begin{equation}
H_{int} (t) = {1 \over 2 V} \sum_{\k} \nu (\k) N_{\k} (t)  n_{-\k} (t)
\label{2.7}
\end{equation}
and here
\begin{eqnarray}
N_\k (t) &=& \sum_{\q} a^{\dag}_{\q} a_{\q + \k} \ e^{ i (E_\q - E_{ \q + \k})t}
\\
n_\k  (t)  &=& \sum_{\q} b^{\dag}_{\q} b_{\q + \k} \ e^{ i (\omega_{\q} - \omega_{\q + \k} ) t }
\label{2.8}
\end{eqnarray}

It is enlightening to look at a simple scattering situation to determine
how the environment stores information about the system (which in turn
determines the preferred basis).
Suppose, for simplicity, that
the distinguished system is classical, and consider what happens when the environment
scatters off it. Suppose the environment starts in an initial momentum
state $ | \k_0 \rangle $ and scatters into a final state $ | \k_f \rangle$.
The scattering amplitude for this process, to first order, is
\begin{eqnarray}
\langle \k_f | S | \k_0 \rangle  &=& { i \over 2 V}  \int_{ - \infty}^{\infty}
dt \sum_{\k} \nu (\k) N_{\k} (t)
\langle \k_f | n_{-\k} (t) | \k_0 \rangle
\nonumber \\
&=& {i \over 2 V} \ \nu (\k ) \ \int dt \  \ N_{\k} (t) \ e^{ i (\omega_{\k_f} -
\omega_{\k_0} ) t }
\label{2.9}
\end{eqnarray}
where $ \k = \k_f - \k_0 $. This simple result shows that a single scattering event by the
environment stores information about the Fourier transform (in space and time) of the number
density. It is in this sense that the number density has a preferred status -- this is
the variable that is measured most directly by the environment. (An analagous
result holds in linear oscillator models \cite{Hal1}).

The measured variables above are of course non-local in time, involving a temporal Fourier
transform of the number density, so cannot in fact be compatible with a Markovian
master equation of the Lindblad form.
Under a reasonable slow motion assumption, the system timescale
is much slower than the environment
timescale, and we may ignore the time-dependence in $ N_{\k} (t)$, yielding
\begin{equation}
\langle \k_f | S | \k_0 \rangle
= {i \over 2 V} \ \nu (\k ) \   \ N_{\k} \ \delta (\omega_{\k_f} - \omega_{\k_0} )
\label{2.10}
\end{equation}
This corresponds more directly to a Markovian master equation, as we shall see.

\section{Derivation of the Master Equation}

Following the method first used by Joos and Zeh \cite{JoZ}, we may derive
the master equation for the reduced density operator $\rho$ of the system
by considering the scattering of the environment off the system, to second
order in interactions. We assume that the system and environment are initally
uncorrelated, so the total density operator is
\begin{equation}
\rho_T = \rho_0 \otimes \rho_{\E}
\label{3.1}
\end{equation}
We also assume that each scattering event takes place on a timescale
which is extremely short compared to the timescale of system dynamics. This means
that in an interval of time $\Delta t $ which is long for the environment
but short for the system, we may write,
\begin{equation}
\rho_T (t + \Delta t ) = S \rho_T (t) S^{\dag}
\label{3.2}
\end{equation}
(where we are using the interaction picture).
Expanding (\ref{3.2}) to second order, the $S$-matrix may be
written,
\begin{equation}
S = 1 + i U_1 - U_2
\label{3.3}
\end{equation}
where
\begin{equation}
U_1 =   - \int_{ - \infty}^{\infty} dt \ H_{int} (t)
\label{3.4}
\end{equation}
and
\begin{equation}
U_2 = \half \int dt_1 \int dt_2 \ {\rm T} \left( H_{int} (t_1) H_{int} (t_2) \right)
\label{3.5}
\end{equation}
The requirement of unitarity, $ S^{-1} = S^{\dag}$, implies that $ U_1 = U_1^{\dag}$
and
\begin{equation}
U_2 + U_2^{\dag} = U_1^2
\label{3.6}
\end{equation}
We will therefore write
\begin{equation}
U_2 = \half U_1^2 + i B
\label{3.7}
\end{equation}
where $B = B^{\dag}$, so we now have
\begin{equation}
S = 1 + i (U_1 - B) - \half U_1^2
\label{3.8}
\end{equation}
Inserting this in (\ref{3.2}), we obtain
\begin{equation}
{ d \rho_T \over dt} \ \Delta t = i [ U_1 - B , \rho_T ]
+ U_1 \rho_T U_1 - \half U_1^2 \rho_T - \half \rho_T U_1^2
\label{3.9}
\end{equation}
We now trace Eq.(\ref{3.9}) over the environment to obtain the master
equation for the system density operator $\rho$. As is usual in this
sort of model, we assume that the environment is so large that
its state is essentially unaffected by the interaction with the system.
Since the total density operator starts out in the factored
state (\ref{3.1}), this then means that, to a good approximation,
$\rho_T$ persists in the approximately factored form
$ \rho \otimes \rho_{\E}$, and we may insert this
in the right-hand side of Eq.(\ref{3.9}) \cite{Gar}.
We thus obtain the preliminary form for the master equation
\begin{equation}
{ d \rho \over dt} \ \Delta t = i [ {\rm Tr}_\E (U_1 \rho_{\E}) - {\rm Tr}_\E ( B \rho_\E), \rho ]
+ {\rm Tr}_{\E} \left(  U_1 \rho_T U_1 - \half U_1^2 \rho_T - \half \rho_T U_1^2 \right)
\label{3.10}
\end{equation}

We now work out these terms in more detail. We will employ the simple but
useful slow motion approximation, in which we ignore the time-dependence of $N_{\k} (t)$.
(Corrections to this approximation are considered in Ref.\cite{HaDo}).
This implies that
\begin{equation}
U_1 \approx  - { 1 \over 2 V} \sum_{\k} \nu (\k) N_{\k}
\ \sum_{\q} b^{\dag}_{\q} b_{\q - \k} \  2 \pi   \delta (\omega_{\q} - \omega_{\q - \k} )
\label{3.11}
\end{equation}
The important terms for decoherence are the final three terms on the right-hand
side of (\ref{3.11}). When traced, these give,
\begin{equation}
{\rm Tr}_{\E} \left( U_1 \rho_T U_1 - \half U_1^2 \rho_T - \half \rho_T U_1^2 \right ) =
\sum_{\k \k'} c(\k, \k') \left( N_{\k'}  \rho N_{\k}
- \half N_{\k} N_{\k'} \rho - \half \rho N_{\k} N_{\k'} \right)
\label{3.12}
\end{equation}
where
\begin{equation}
c(\k, \k') = \nu (\k) \nu (\k')
\sum_{\q \q'}
 \ \delta (\omega_{\q} - \omega_{\q - \k} )  \delta (\omega_{\q'} - \omega_{\q' - \k' } )
\ \langle b^{\dag}_{\q} b_{\q - \k}  b^{\dag}_{\q'} b_{\q' - \k'} \rangle_{\E}
\label{3.13}
\end{equation}
We will take the environment to be a thermal state, which is diagonal in the
momentum states. It follows that
\begin{equation}
\langle b^{\dag}_{\q} b_{\q - \k}  b^{\dag}_{\q'} b_{\q' - \k'} \rangle_{\E}
\propto \delta_{\q, \q' - \k'} \ \delta_{\q', \q- \k}
\label{3.14}
\end{equation}
This implies $\k = - \k'$, and also that the two delta-functions are the same
in Eq.(\ref{3.13}).
We then interpret the square of the delta-function in the usual way,
\begin{eqnarray}
\left[ \delta (\omega_{\q} - \omega_{\q - \k} ) \right]^2 &=& \delta (0)
\ \delta (\omega_{\q} - \omega_{\q - \k} )
\nonumber \\
&=& \frac {\Delta t } {2 \pi}  \ \delta (\omega_{\q} - \omega_{\q - \k} )
\label{3.14}
\end{eqnarray}
We now have
\begin{equation}
c(\k, \k') = \delta_{\k, - \k'} \ c(\k) \frac{\Delta t} {2 \pi}
\label{3.14b}
\end{equation}
where
\begin{eqnarray}
c(\k ) &=&  \frac {1} {2 \pi} \left | \nu( \k) \right|^2 \sum_\q \   \ \delta (\omega_{\q} - \omega_{\q - \k} )
\langle b^{\dag}_{\q} b_{\q - \k}  b^{\dag}_{\q- \k} b_{\q} \rangle_{\E}
\nonumber \\
&=& \frac {1} {2 \pi} \left | \nu( \k) \right|^2 \sum_\q \   \ \delta (\omega_{\q} - \omega_{\q - \k} )
\langle b^{\dag}_{\q} b_{\q} \rangle_{\E} \left( \langle b^{\dag}_{\q - \k} b_{\q- \k } \rangle_{\E}
+ 1 \right)
\label{3.15}
\end{eqnarray}
The terms involving environment averages have the usual thermal form (for a bosonic environment),
\begin{equation}
\langle b^{\dag}_{\q} b_{\q} \rangle_{\E} = \frac {1} { e^{\beta (\omega_\q - \mu) } - 1 }
\end{equation}
where $\b = 1/ k T $ with $T$ temperature, and $\mu$ is the chemical potential.

The form Eq.(\ref{3.14b}) means that the important terms in the master equation are of the Lindblad form,
\begin{equation}
{ \rm Tr}_{\E} \left( U_1 \rho_T U_1 - \half U_1^2 \rho_T - \half \rho_T U_1^2 \right ) =
\Delta t \ \sum_{\k } c(\k ) \left( N_{\k}  \rho N^{\dag}_{\k}
- \half N^{\dag}_{\k} N_{\k} \rho - \half \rho N^{\dag}_{\k} N_{\k} \right)
\label{3.16}
\end{equation}
where we have used the fact that $ N^{\dag}_{\k} = N_{-\k} $.
The remaining two terms in Eq.(\ref{3.10}) clearly just modify the unitary
dynamics of the system. First we have
\begin{equation}
{\rm Tr}_\E \left( U_1 \rho_\E \right) = { 1 \over 2 V} \sum_{\k} \nu (\k) N_{\k}
\ \sum_{\q} \langle b^{\dag}_{\q} b_{\q - \k} \rangle_{\E} \  2 \pi   \delta (\omega_{\q}
- \omega_{\q - \k} )
\label{3.17}
\end{equation}
Clearly from the term $ \langle b^{\dag}_{\q} b_{\q - \k} \rangle_{\E} $ this
expression will be zero unless $\k = 0$, and therefore it is proportional to
$N$, the total particle number operator (although the overall coefficient will need
to be regularized). This therefore contributes a term to the master equation
of the form $ [ N, \rho ] $. We assume that there is a fixed number of system
particles so it is reasonable to take this term to be zero.

The other remaining term in Eq.(\ref{3.10}) involves the time ordering terms in $U_2$ and is a
bit more complicated
to evaluate. Fortunately, the detailed form of this expression is not needed here,
and it can in fact be easily shown that this term has the form
\begin{equation}
{\rm Tr}_\E ( B \rho_\E) = \Delta t \sum_{\k} \ d(\k) \ N_{\k} N^{\dag}_{\k}
\label{3.18}
\end{equation}
for some coefficient $d (\k) $ which we will not need.
Inserting all these
results in Eq.(\ref{3.10}), the factors of $\Delta t$ all drop out, and we
obtain, in the Schr\"odinger picture,
\begin{equation}
{ d \rho \over d t } = -i [ H_0 - \sum_{\k} \ d(\k)  \ N_{\k} N^{\dag}_{\k}, \rho]
+ \sum_{\k } c(\k ) \left( N_{\k}  \rho N^{\dag}_{\k}
- \half N^{\dag}_{\k} N_{\k} \rho - \half \rho N^{\dag}_{\k} N_{\k} \right)
\label{3.19}
\end{equation}
As desired, this is the Lindblad form with the Lindblad operators given
by
\begin{equation}
L_{\k} = c^{\half} (\k) N_{\k}
\end{equation}
We have therefore produced a derivation of the master equation for a scattering
environment which shows very clearly the key role of local number
density as the preferred basis,
as indicated by the simple scattering calculation, Eq.(\ref{2.10}).

It is interesting to note that the decoherence effect is second order in
interactions, but we were able to anticipate it from the simple first
order calculation, Eq.(\ref{2.10}). The reason for this is the relationship
Eq.(\ref{3.7}), which shows that the important part of the second order
terms is the square of the first order terms, and this is a consequence of unitarity.

\section{Comparison with Previous Works}

It is now important to check that the master equation we have derived reproduces
known results when we restrict to the one-particle sector for the system, where
a number of derivations have been given in a quantum mechanical framework
\cite{JoZ,GaF,Dio,HoSi,Vac}.
In the one-particle sector we may work with a density matrix $\rho (\k, \k')
= \langle \k | \rho | \k' \rangle$,
or equivalently $\rho (\x, \y) $ in the position representation.
We use the relations
\begin{eqnarray}
\left[ N_{\mathbf{q}},a_{\mathbf{k}}^{\dagger} \right]
& = & a_{\mathbf{k-q}}^{\dagger}
\\
\left[ N_{\mathbf{q}},a_{\mathbf{k}} \right]
& = & -a_{\mathbf{k+q}}
\end{eqnarray}
These relations imply that
\begin{eqnarray}
N_{\mathbf{q}}\rho(\mathbf{k},\mathbf{k'}) N_{\mathbf{-q}}
&=& \rho(\mathbf{k-q},\mathbf{k'-q})
\\
N_{\mathbf{-q}}N_{\mathbf{q}}\rho(\mathbf{k},\mathbf{k'})
&=&\rho(\mathbf{k},\mathbf{k'})
\\
\rho(\mathbf{k},\mathbf{k'})N_{\mathbf{-q}}N_{\mathbf{q}}
&=& \rho(\mathbf{k},\mathbf{k'})
\end{eqnarray}
In the position representation, this means
\begin{equation}
N_{\mathbf{k}}\rho(\mathbf{x},\mathbf{y}) N_{\mathbf{-k}}=
e^{i\mathbf{k} \cdot (\mathbf{x}-\mathbf{y})}\rho(\mathbf{x},\mathbf{y})
\end{equation}
The master equation for the one-particle density operator $\rho (\x, \y )$ is
then
\begin{equation}
\frac { \partial \rho (\x, \y ) } {\partial t}
= - i \langle \x | \left[ H_0, \rho \right] | \y \rangle
- F (\x - \y ) \rho (\x, \y)
\label{6.7}
\end{equation}
where
\begin{equation}
F( \x - \y) = \frac{1} {(2 \pi)^5} \int d^3q d^3k \ \left| \nu
({\k}) \right|^2 \ n_{\mathbf{q}} (n_{\mathbf{q-k}}+1)\ \delta
(\omega_{\mathbf{q}}- \omega_{\mathbf{q-k}}) \ (1- e^{i\mathbf{k}
\cdot(\mathbf{x-y})}) \label{4.1}
\end{equation}
Note that the term involving the coefficient $d (\k) $ in Eq.(\ref{3.19}) drops
out because $ [ N_{\k} N^{\dag}_{\k} , \rho ] = 0$ in the one-particle sector.

Eq.(\ref{6.7}) is of the same general form as earlier results \cite{JoZ,GaF,Dio}.
To compare in detail, we first introduce
the quantity
\begin{equation}
f(\mathbf{k},\mathbf{k'})=\frac{ m } {2 \pi }
\nu ( \k - \k')
\end{equation}
(which appears in the usual Born approximation to first order scattering).
Then, letting $ \k \rightarrow -\k + \q $ in (\ref{4.1}), we get
\begin{equation}
F( \r) = \frac{ 1 } { (2 \pi)^3 m^2}
\int d^3q d^3k
\ \left| f (\q, \k) \right|^2 \ n_{\mathbf{q}}
(n_{\mathbf{k}}+1)\ \delta (\omega_{\mathbf{q}}- \omega_{\mathbf{k}})
\ (1- e^{i(\mathbf{q-k})\cdot \mathbf{r}})
\label{4.2}
\end{equation}
The delta-function implies that $\q^2 = \k^2 $, and we find that
\begin{equation}
F( \r ) = \frac{ 1 } { (2 \pi)^3 m^2}
\int dq \ q^3 \ n_q ( n_q + 1 ) \ \int d\Omega d\Omega'
|f(\mathbf{q},\mathbf{k})|^2(1- e^{i (\mathbf{q-k}) \cdot \mathbf{r} })
\end{equation}
For $ n_q \ll 1 $, and identifying
$ (1/ 2 \pi^2) q^2 n_q dq $ as the fraction of particles
with momentum magnitude between $q$ and $q + dq$, we find agreement
with the careful derivation of Hornberger and Sipe \cite{HoSi},
which in turn agrees with recent experiments which measure the
decoherence rate \cite{Exp}. (Hornberger and Sipe corrected erroneous numerical
factors in some of the earlier derivations \cite{JoZ,GaF}, but the
qualitative and order of magnitude predictions of these earlier
works are correct. See also the cross-check of Adler \cite{Adl}).

\section{Acknowledgements}

I am grateful to Peter Dodd, Lajos Diosi and Erich Joos for useful conversations.

\bibliography{apssamp}

\begin{thebibliography}{10}



\bibitem{GRW} G.C.Ghirardi, A.Rimini and T.Weber, Phys. Rev. D 34,
470-491 (1986).

\bibitem{JoZ} E.Joos and H.D.Zeh,  Z.Phys. {B59}, 223 (1985).

\bibitem{GaF} M.R.Gallis and G.N.Fleming, Phys. Rev. {A42}, 38-48 (1990).


\bibitem{Dio} L.Diosi, Europhys. Lett. {30}, 63 (1995).

\bibitem{Vac} B.Vacchini, Int. J. Theor. Phys. 44 (2005) 1011-1021

\bibitem{HoSi} K.Hornberger and J.E.Sipe,  Phys. Rev. A 68, 012105 (2003)


\bibitem{Omn1} R.Omn\`es, Phys.Rev. {A56}, 3383 (1997).



\bibitem{CaL} A.Caldeira and A.Leggett, Physica {121A}, 587 (1983).

\bibitem{HPZ} B.L.Hu, J.Paz and Y. Zhang, Phys.Rev. {D45}, 2843(1992); {
D47}, 1576(1993).

\bibitem{HaYu} J.J.Halliwell and T.Yu, Phys.Rev. {53}, 2012 (1996).


\bibitem{HaDo} P.Dodd and J.J.Halliwell, Phys Rev D67, 105018 (2003).


\bibitem{UnZu} W.G.Unruh and W.H.Zurek, Phys.Rev. {D40},
1071 (1989).

\bibitem{AnZo} C.Anastopoulos and A.Zoupas, Phys.Rev.
{D58}, 105006 (1998).

\bibitem{QBM}
I.R.Senitzky, Phys.Rev. {119}, 670 (1960);
J.Schwinger, J.Math.Phys. {2}, 407 (1961);
G.W.Ford, M.Kac and P.Mazur, J.Math.Phys. {6}, 504
(1965);
G.S.Agarwal,
Phys.Rev. { A3}, 828 (1971); Phys.Rev. {A4}, 739 (1971);
V.Hakim and V.Ambegaokar, Phys.Rev. {A32}, 423
(1985);
H.Dekker, Phys.Rev. {A16}, 2116 (1977); Phys.Rep. {80}, 1
(1991);
H.Grabert, P.Schramm, G-L. Ingold, Phys.Rep. {168},
115 (1988).



\bibitem{Lin} G.Lindblad, Comm.Math.Phys. {48}, 119 (1976).


\bibitem{Dio2} L.Di\'osi, Europhys.Lett. {22}, 1 (1993).

\bibitem{HaZ} J.J.Halliwell and A.Zoupas,  Phys.Rev.
{ D52}, 7294 (1995); {D55}, 4697 (1997).


\bibitem{Har6} J.B.Hartle, in, {\it Proceedings of
the Cornelius Lanczos International Centenary Confererence},
edited by J.D.Brown, M.T.Chu, D.C.Ellison and R.J.Plemmons
(SIAM, Philadelphia, 1994)
also available as e-print gr-qc/9404017 (1994).


\bibitem{Zur1} See for example, W. Zurek, Physics Today {40}, 36 (1991)

\bibitem{Hal0} J.J.Halliwell, Contemporary Physics 46, 93 (2005).

\bibitem{Exp} K.Hornberger, S.Uttenthaler, B.Brezger,
L.Hackermuller, M.Arndt and A.Zeilinger, Phys. Rev. Lett.
90, 160401 (2003).

\bibitem{Hor} K.Hornberger, e-print quant-ph/0607085.

\bibitem{Adl} S.Adler, e-print quant-ph/0607109.


\bibitem{Joos} E. Joos, H.D. Zeh, C. Kiefer, D. Giulini, J. Kupsch
and I.O. Stamatescu, {\it Decoherence and the Appearance of a Classical
World in Quantum Theory}  (Springer-Verlag, 2003).


\bibitem{DGHP} L.Di\'osi, N.Gisin, J.Halliwell and I.C.Percival,
Phys.Rev.Lett. {74}, 203 (1995).


\bibitem{HalHydro} J.J.Halliwell, Phys.Rev.Lett. 83 (1999) 2481;
Phys Rev D68, 025018 (2003).


\bibitem{GeH1} M.Gell-Mann and J.B.Hartle, Phys.Rev. {D47},
3345 (1993).

\bibitem{GeH2} M.Gell-Mann and J.B.Hartle, in {\it Complexity, Entropy
and the Physics of Information, SFI Studies in the Sciences of
Complexity}, Vol. VIII, W. Zurek (ed.) (Addison Wesley, Reading,
1990); and in {\it Proceedings of the Third International
Symposium on the Foundations of Quantum Mechanics in the Light of
New Technology}, S. Kobayashi, H. Ezawa, Y. Murayama and S. Nomura
(eds.) (Physical Society of Japan, Tokyo, 1990).



\bibitem{Har1} J.B.Hartle, in {\it Proceedings of the 1992 Les Houches Summer
School, Gravitation et Quantifications}, edited by B.Julia and
J.Zinn-Justin (Elsevier Science B.V., 1995).


\bibitem{Gri} R.B.Griffiths, J.Stat.Phys. {36}, 219 (1984);
{ Phys.Rev.Lett.} { 70}, 2201 (1993).

\bibitem{Omn} R. Omn\`es, J.Stat.Phys. {53}, 893 (1988);
{53}, 933 (1988); {53}, 957 (1988); {57}, 357 (1989);
Ann.Phys. { 201}, 354 (1990); Rev.Mod.Phys. {64}, 339
(1992).

\bibitem{Wig} N.Balazs and B.K.Jennings, Phys.Rep. {104}, 347 (1984),
M.Hillery, R.F.O'Connell, M.O.Scully and E.P.Wigner, Phys.Rep. {106}, 121 (1984);
V.I.Tatarskii, {Sov.Phys.Usp} {26}, 311 (1983).


\bibitem{Amb} V.Ambegaokar, Ber.Bunsenges.Phys.Chem. {95},
400 (1991). (This author in turn cites a private communication
from P.Pechukas.)

\bibitem{AnHa} C.Anastopoulos and J.J.Halliwell, Phys.Rev. D51, 6870 (1995).

\bibitem{Zub} D.N.Zubarev, {\it Nonequilibrium Statistical
Thermodynamics} (Consultants Bureau, New York, 1974).

\bibitem{FeWa} A.L.Fetter and J.D.Walecka, {\it Quantum Theory
of Many-Particle Systems} (McGraw-Hill, 1971).


\bibitem{Hal1} J.J.Halliwell, Phys.Rev. {D60}, 105031
(1999).

\bibitem{Gar} C.W.Gardiner, {\it Quantum Noise} (Springer-Verlag,
Berlin, 1991).






\end{thebibliography}

\end{document}